\documentclass[aps,twocolumn,showpacs]{revtex4}
\usepackage{dcolumn}
\usepackage{graphicx}
\usepackage{amsmath}
\usepackage{amsfonts}
\usepackage{amssymb}
\usepackage{psfrag}
\usepackage{wrapfig}
\usepackage{subfigure}
\usepackage{makeidx}
\usepackage{bm}
\usepackage{epsf}

\begin{document}

\title{High-order Rogue Waves in Vector Nonlinear Schr\"{o}dinger Equations}
\author{Liming Ling$^{1}$, Boling Guo$^{2}$, }
\author{Li-Chen Zhao$^{3}$}\email{zhaolichen3@163.com}
\address{$^1$Department of Mathematics, South China University of Technology,
Guangzhou 510640, China}
\address{$^2$Institute of Applied Physics and Computational Mathematics,
Beijing 100088, China}
\address{$^3$Department of Physics, Northeast University, Xi'an 710069, China}

\date{November 6, 2013}
\begin{abstract}
We study on dynamics of high-order rogue wave in two-component coupled nonlinear Schr\"{o}dinger equations. We find four fundamental rogue waves can emerge for second-order vector RW in the coupled system, in contrast to the high-order ones in single component systems. The distribution shape can be  quadrilateral, triangle, and line structures through varying the proper initial excitations given by the exact analytical solutions. Moreover, six fundamental rogue wave can emerge on the distribution for second-order vector rogue wave, which is similar to the scalar third-order ones. The distribution patten for vector ones are much abundant than the ones for scalar rogue waves. The results could be of interest in such diverse fields as Bose-Einstein
condensates, nonlinear fibers, and superfluids.

\end{abstract}

 \maketitle

\emph{Introduction}--- Rogue wave (RW) is the name given by
oceanographers to isolated large amplitude wave, which occurs more
frequently then expected for normal, Gaussian distributed,
statistical events \cite{Onorato,Kharif1,Osborne}. It depicts a
unique event that seems to appear from nowhere and disappear without
a trace, and can appear in a variety of different contexts
\cite{Ruban,N.Akhmediev,Kharif,Pelinovsky}. RW has been observed
experimentally in nonlinear optics by Solli group and Kibler group
\cite{Solli, Kibler}, water wave tank by Chabchoub group
\cite{Chabchoub}, and even in plasma system by Bailung group
\cite{Bailung}.  These experimental studies suggest that the
rational solutions of related dynamics equations can be used to
describe these RW phenomena \cite{R.Osborne,N.A}. Moreover, there
are many different pattern structures for high-order RW
\cite{Yang,Ling,He}, which can be understood as a nonlinear
superposition of fundamental RW (the first order RW). Recently, many
efforts are devoted to classify the hierarchy for each order RW
\cite{Ling2,Akhmediev}, since these superpositions are nontrivial
and admit only a fixed number of elementary RWs in each high order
solution (3 or 6 fundamental ones for the second order or third
order one).

Recent studies are extended to
 RWs in multi-component coupled systems(vector RW), since a variety of complex systems, such as Bose-Einstein condensates,
nonlinear optical fibers, etc., usually involve more than one
component \cite{Becker,Engels,Tang}. For the coupled system, the usual coupled effects are cross-phase modulation terms. The linear stability analyzes indicate that the cross-phase modulation term can vary the instability regime characters \cite{Forest,Wright,Chow}. Moreover, for scalar system, the velocity of the background field has no real effects on pattern structure for nonlinear localized waves, since the corresponding solutions can be correlated though trivial Galileo transformation. But the relative velocity between different component field has real physical effects, and can not be erased by any trivial transformation. Therefore, the extended studies on vector ones are not trivial.
In the two-component coupled
systems,  dark RW was presented numerically \cite{Bludov} and analytically
\cite{zhao2}.  The interaction between RW and other nonlinear waves is
also a hot topic of great interest \cite{Baronio,Ling3,zhao2}. For example, it
has been shown that RW attracts dark-bright wave. In three-component
coupled system, four-petaled flower structure RW was reported very recently in \cite{Zhao3,Degasperis}. These studies indicate that there are much abundant pattern dynamics for RW in the multi-component coupled systems, which are quite distinctive from the ones in scalar systems.

After the Peregrine RW has been observed experimentally in many
different physical systems, high-order RW were excited successfully
in a water wave tank, mainly including the RW triplets for second
order one \cite{Chabchoub2} and up to the fifth-order ones
\cite{Chabchoub3}. This suggests that high order analytic RW
solution is meaningful physically and can be realized experimentally
\cite{Erkintalo}. However, as far as we know, the high order vector
RW has not been taken seriously until now. As high order scalar RWs
are nontrivial superpositions of elementary RWs, the high order
vector ones could be nontrivial and possess more abundant dynamics
characters.  The knowledge about them would enrich our realization
and understanding of RW's complex dynamics.

In this letter,  we
introduce a family of high-order rational solution in two-component
nonlinear Schr\"{o}dinger equations(NLSE), which describe RW
phenomena in multi-component systems prototypically. We find that
four fundamental RWs can emerge for the second-order vector RW in the
coupled system, which is quite different from the scalar high order
ones for which it is impossible for four fundamental RWs to emerge.
Moreover, six fundamental rogue wave can emerge on the distribution
for second order vector rogue wave. The number of RW is identical to
the one for scalar third order RW. But their distribution patten on
the temporal-spatial distribution can be quite distinctive.

 \emph{The two-component coupled model}---
 We
begin with the well known two-coupled NLSE in dimensionless form
\begin{equation}\label{cnls}
    \begin{split}
      i q_{1,t}+q_{1,xx}+2(|q_1|^2+|q_2|^2)q_1 &=0,  \\
      i q_{2,t}+q_{2,xx}+2(|q_1|^2+|q_2|^2)q_2 &=0,
    \end{split}
\end{equation}
which admits the following Lax pair:
\begin{equation}\label{Lax}
    \begin{split}
       \Phi_x &=(i\lambda \Lambda+Q)\Phi,\\
       \Phi_t &=(3i\lambda^2\Lambda+3\lambda Q+i\sigma_3(Q_x-Q^2))\Phi,
    \end{split}
\end{equation}
where
\begin{equation*}
    Q=\begin{pmatrix}
        0 & q_1 & q_2 \\
        -\bar{q}_1 & 0 & 0 \\
        -\bar{q}_2 & 0 & 0 \\
      \end{pmatrix}
    ,\,\left.
         \begin{array}{ll}
          \Lambda&=\mathrm{diag}(-2,1,1),  \\
           \sigma_3&=\mathrm{diag}(1,-1,-1),
         \end{array}
       \right.
\end{equation*}
the symbol overbar represents complex conjugation.
The compatibility condition $\Phi_{xt}=\Phi_{tx}$ gives the CNLS \eqref{cnls}.

We can convert the system \eqref{Lax} into a new linear system
\begin{equation}\label{newlax}
    \begin{split}
   \Phi[1]_x &=(i\lambda \Lambda+Q[1])\Phi[1],\\
       \Phi[1]_t &=(3i\lambda^2\Lambda+3\lambda Q[1]+i\sigma_3(Q[1]_x-Q[1]^2))\Phi[1],
        \end{split}
\end{equation}
 by the following elementary
Darboux transformation
\begin{equation}\label{DT}\begin{split}
                             \Phi[1]&=T\Phi,\,T=I+\frac{\bar{\lambda}_1-\lambda_1}{\lambda-\bar{\lambda}_1}\frac{\Phi_1\Phi_1^{\dag}}{\Phi_1^{\dag}\Phi_1}, \\
                              Q[1]&=Q+i(\bar{\lambda}_1-\lambda_1)[P_1,\Lambda],
                          \end{split}
\end{equation}
where $\Phi_1$ is a special solution for system \eqref{cnls} at
$\lambda=\lambda_1$, the symbol $^{\dag}$ represents the hermitian
transpose.
 It is well known that the standard N-fold Darboux transformation(DT) should be done with different spectrum parameters, or there will be some singularity in the DT matrix. The generalized DT was presented to solve this problem in \cite{Ling}, which can  be used to derive high-order RW conveniently by taking a
special limit about the parameter $\lambda$.

The studies on first order vector RW in the ref. \cite{Ling3,zhao2},
indicate that there should be some restriction condition on the
background fields to obtain the general RW solutions for CNLS. In
ref. \cite{Ling3}, they choose the seed solution
\begin{equation*}
    q_1=\alpha e^{i\left
    (\frac{1}{2}\alpha x+\frac{15}{4}\alpha^2t\right)},\,\, q_2=\alpha e^{i\left
    (-\frac{1}{2}\alpha x+\frac{15}{4}\alpha^2t\right)}.
\end{equation*}
The wave vector difference between the two components should satisfy certain relation with the background amplitudes and nonlinear coefficients \cite{zhao2}, we choose a more simplified form to derive RW solution. In fact, the parameter $\alpha$ can be re-scaled by scaling transformation $$(q_1(x,t),q_2(x,t))\rightarrow (\alpha q_1(\alpha x,\alpha^2 t),\alpha q_2(\alpha x,\alpha^2 t)).$$ Thus we can consider a simple seed solution as the background where RW exist without losing generality
\begin{equation}\label{seed}
    q_1=\exp{\theta_1},\,\,q_2=\exp{\theta_2},
\end{equation}
where $\theta_1=\left [i\left
    (\frac{1}{2}x+\frac{15}{4}t\right)\right]$, $\theta_2=\left[i\left
    (-\frac{1}{2}x+\frac{15}{4}t\right)\right]$.

We have proved that high order RW can be derived by taking a certain
limit of the spectrum parameter \cite{Ling}. To take the limit
conveniently, we set
$\lambda_j=\frac{\sqrt{3}i}{2}(1+\epsilon_j^3)$, $j=1,2,\cdots,N$.
Substituting seed solution \eqref{seed} into equation \eqref{Lax},
we can obtain the fundamental solution
\begin{equation*}
    \Phi_i(\lambda_j)=D\begin{bmatrix}
             [i(\lambda_j+\frac{1}{2})-\xi_i][i(\lambda_j-\frac{1}{2})-\xi_i]\exp{\omega_i}
             \\[4pt]
             [i(\lambda_j-\frac{1}{2})-\xi_i]\exp{\omega_i} \\[4pt]
             [i(\lambda_j+\frac{1}{2})-\xi_i]\exp{\omega_i} \\
           \end{bmatrix}
\end{equation*}
where $i=1,2,3,$
\begin{equation*}
    \begin{split}
      D&=\mathrm{diag}\left(e^{\frac{5it}{2}},\,e^{-\frac{i}{4}(2x+5t)},\,e^{\frac{i}{4}(2x-5t)}\right) \\
      \omega_i&=\xi_ix+\left(i\xi_i^2+2\lambda_j\xi_i+2i\lambda_j^2+\frac{3i}{2}\right)t
    \end{split}
\end{equation*}
and $\xi_i$ satisfy the following cubic equation
\begin{equation}\label{cubic}
    {\xi}^{3}-\left(\frac{9}{2}\,\epsilon_j^{3}+\frac{9}{4}\,\epsilon_j^{6}
 \right) \xi-\frac{3}{2}\,\sqrt {3}\epsilon_j^{3}-\frac{9}{4}\,\sqrt {3}\epsilon_j
^{6}-\frac{3}{4}\,\sqrt {3}\epsilon_j^{9}=0.
\end{equation}
By the Taylor expansion of the fundamental solution form as done in \cite{Ling}, the generalized DT can be used to derive RW solution. However, Taylor expansions of the fundamental solution form are quite complicated which bring much complex calculation. We find this process can be simplified greatly by the following special solution form
\begin{equation}\label{gensol}
   \begin{split}
      \Psi_1(\lambda_j)&=\frac{1}{3}\left(\Phi_1(\lambda_j)+\Phi_2(\lambda_j)+\Phi_3(\lambda_j)\right), \\[4pt]
      \Psi_2(\lambda_j)&=\frac{\sqrt[3]{2}}{3\epsilon_j}\left(\Phi_1(\lambda_j)+\omega^2\Phi_2(\lambda_j)+\omega\Phi_3(\lambda_j)\right),\\[4pt]
      \Psi_3(\lambda_j)&=\frac{\sqrt[3]{4}}{3\epsilon_j^2}\left(\Phi_1(\lambda_j)+\omega\Phi_2(\lambda_j)+\omega^2\Phi_3(\lambda_j)\right),
   \end{split}
\end{equation}
where $\omega=\exp[2i\pi/3]$, which are also the solution of the Lax
pair with the seed solutions \eqref{seed}. We can prove that
\begin{equation}\label{solu}
  \Psi(\epsilon_j)=f\Psi_1+g\Psi_2+h\Psi_3
\end{equation}
where
\begin{equation*}
\begin{split}
  f&=f_1+f_2\epsilon_j^3+f_3\epsilon_j^6+\cdots+f_{N}\epsilon_j^{3(N-1)},\\
  g&=g_1+g_2\epsilon_j^3+g_3\epsilon_j^6+\cdots+g_N\epsilon_j^{3(N-1)},\\
  h&=h_1+h_2\epsilon_j^3+h_3\epsilon_j^6+\cdots+h_N\epsilon_j^{3(N-1)},
\end{split}
\end{equation*}
and $f_i$,$g_i$,$h_i$ are complex numbers, can be expanded around
$\epsilon_j=0$ with the following form
\begin{equation*}
  \Psi(\epsilon_j)
=\Psi^{[1]}+\Psi^{[2]}\epsilon_j^3+\Psi^{[3]}\epsilon_j^6+\cdots+\Psi^{[N]}\epsilon_j^{3(N-1)}+O(\epsilon_j^{3N}).
\end{equation*}
To obtain the vector RW solution, we merely need to take limit
$\epsilon_j\rightarrow 0$ \cite{Ling2}. After performing the
generalized DT, we can present N-th order localized solution on the
plane backgrounds with the same spectral parameter
$\lambda_j=\frac{\sqrt{3}i}{2}(1+\epsilon_j^3)$ as
\begin{equation}\label{NRW}
    \begin{split}
      q_1[N]&=\exp{\theta_1}\frac{\det(M_1)}{\det(M)},  \\
      q_2[N]&=\exp{\theta_2}\frac{\det(M_2)}{\det(M)},
    \end{split}
\end{equation}
where
\begin{eqnarray}
    M_1&=&M-3iY_2^{\dag}Y_1,\,\,M_2=M-3iY_3^{\dag}Y_1,\nonumber\\
     X& =&\begin{bmatrix}
           X_1 \\
           X_2 \\
           X_3 \\
         \end{bmatrix}
=\left[\Psi^{[1]},\,\Psi^{[2]},\,\cdots,\Psi^{[N]}\right],
\end{eqnarray}
$Y_1=X_1e^{-\frac{5it}{2}}$, $Y_2=X_2e^{\frac{i}{4}(2x+5t)}$,
$Y_3=X_3e^{\frac{i}{4}(-2x+5t)}$, and $M=(M_{l,m})_{1\leq
l,m\leq N}$.
The $M_{l,m}$ can be derived by
\begin{equation*}
    \frac{\langle\Psi(\epsilon_j),\Psi(\epsilon_j)\rangle}{\lambda_j-\bar{\lambda}_j}
    =\sum_{l,m=1}^{+\infty,+\infty}M_{l,m}\epsilon_j^{3(m-1)}\bar{\epsilon}_j^{3(l-1)}.
\end{equation*}
\begin{figure}[htb]
\centering
\subfigure[]{\includegraphics[height=36mm,width=42.5mm]{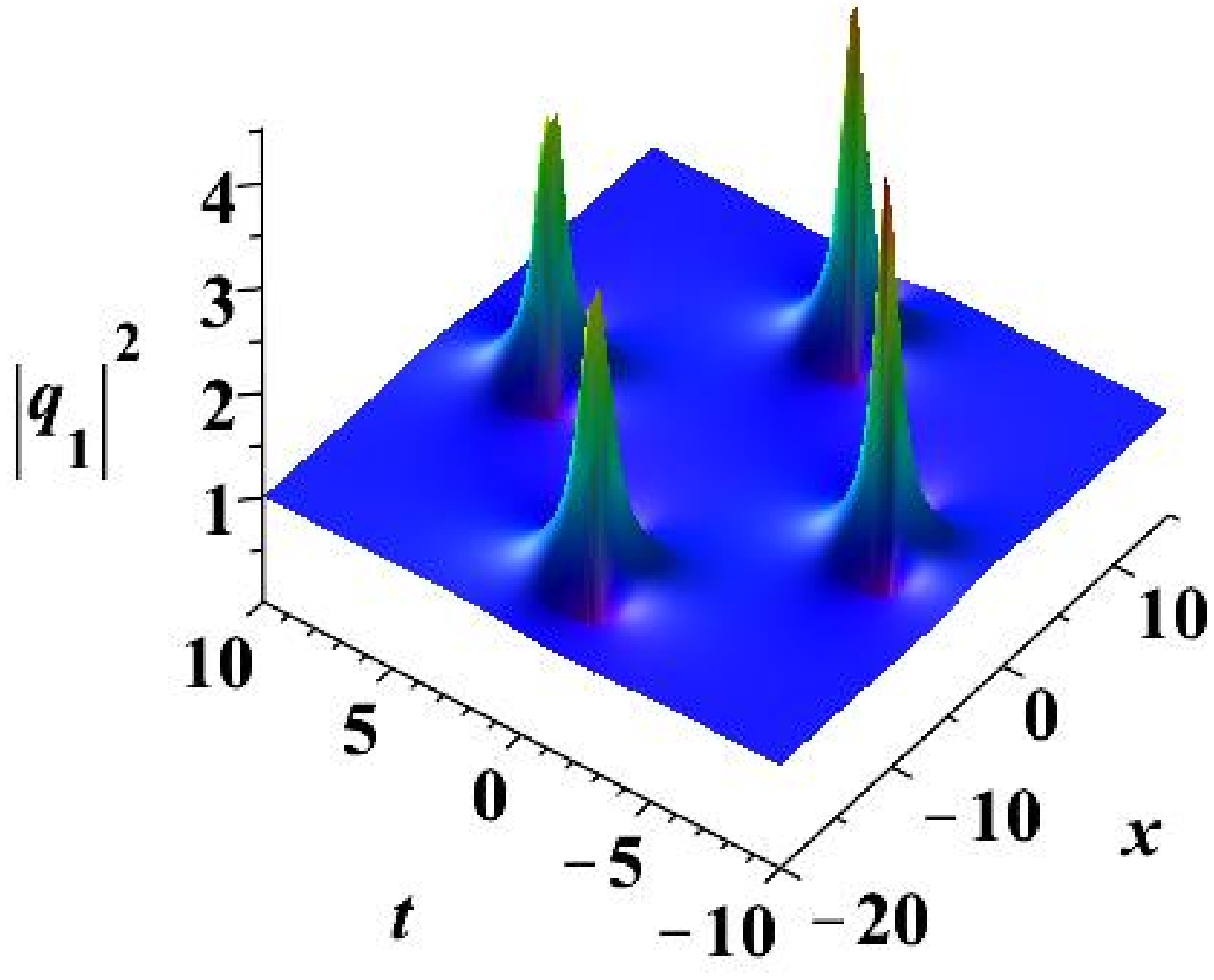}}
\hfil
\subfigure[]{\includegraphics[height=36mm,width=42.5mm]{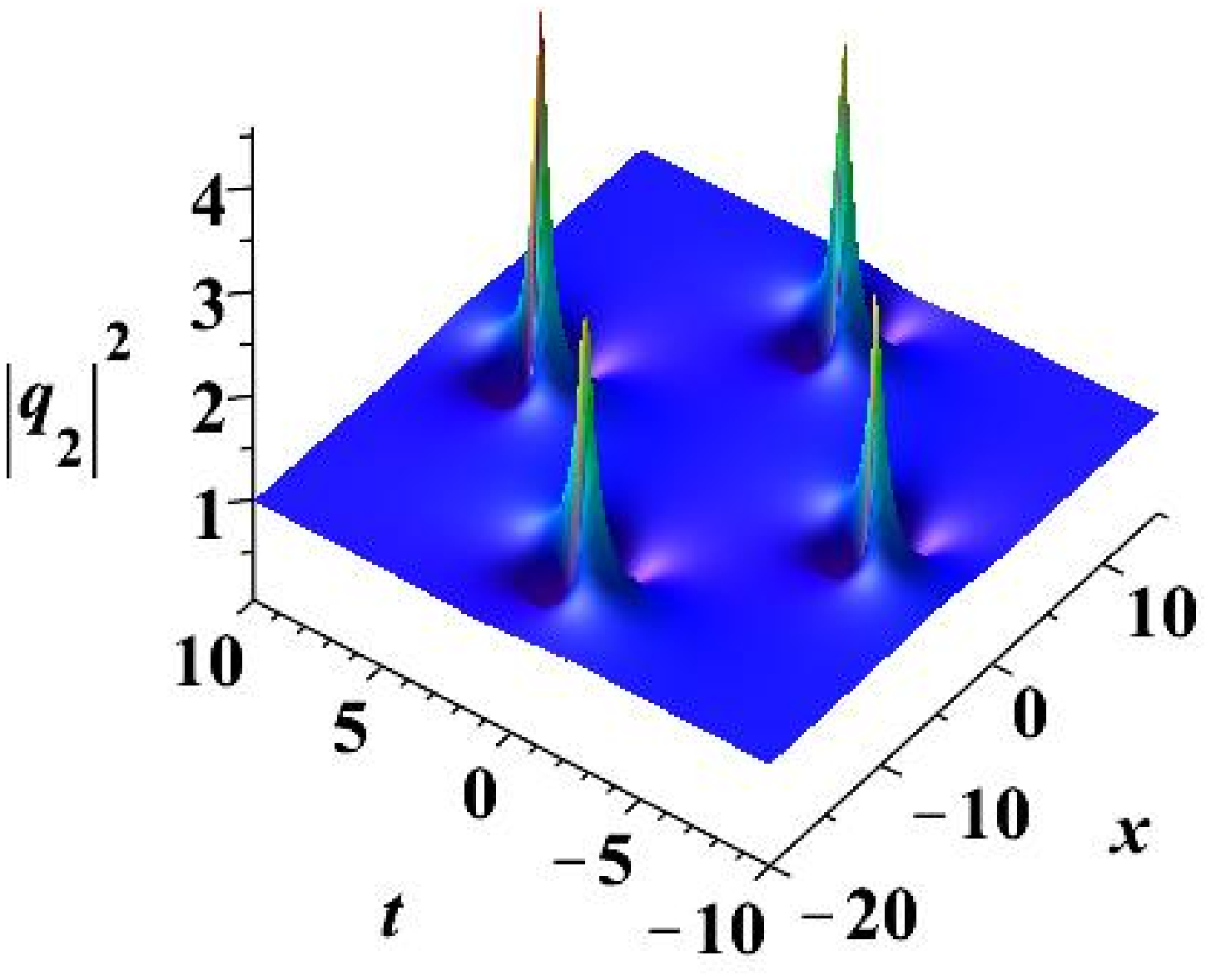}}
\hfil
\subfigure[]{\includegraphics[height=36mm,width=42.5mm]{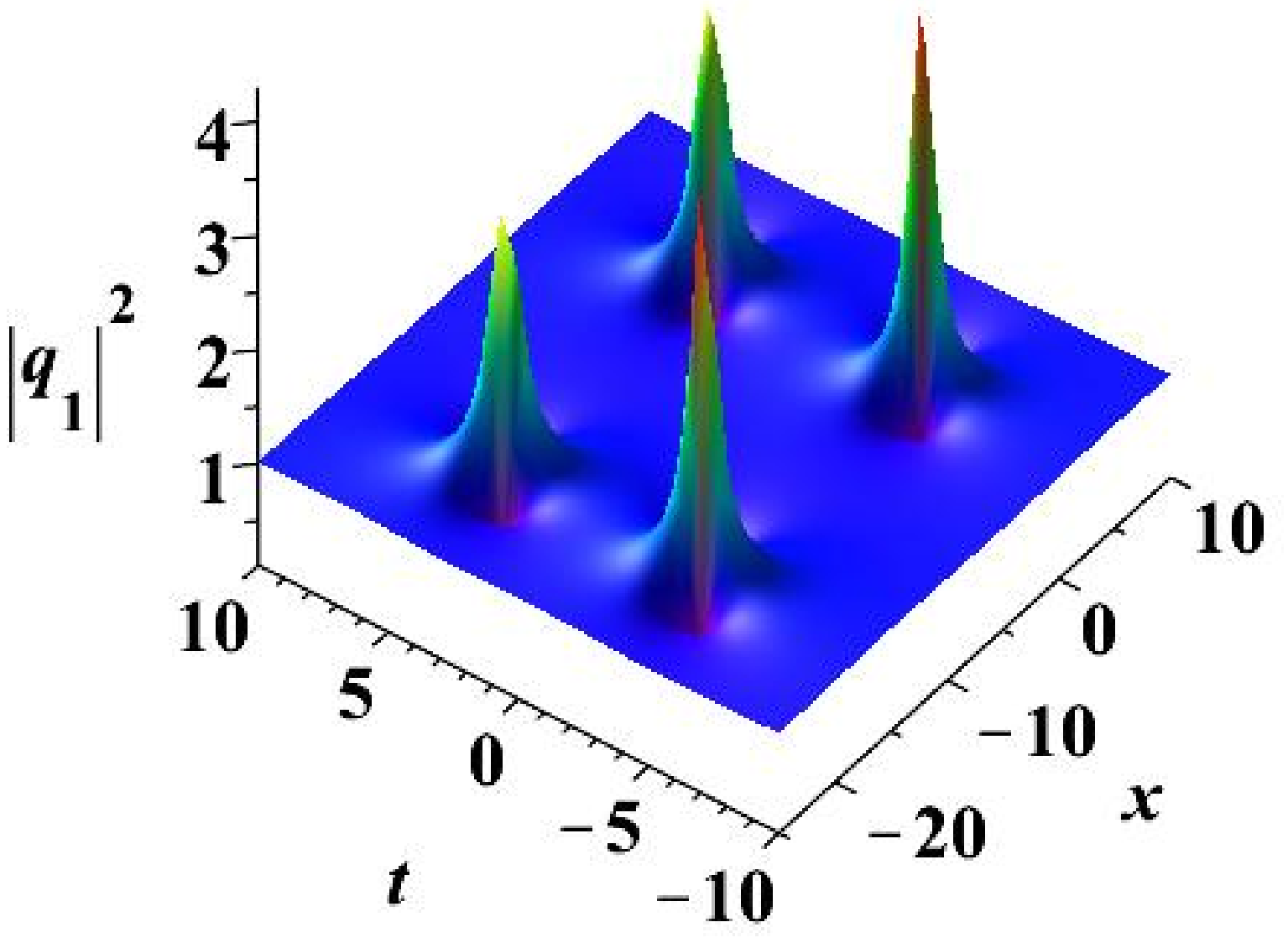}}
\hfil
\subfigure[]{\includegraphics[height=36mm,width=42.5mm]{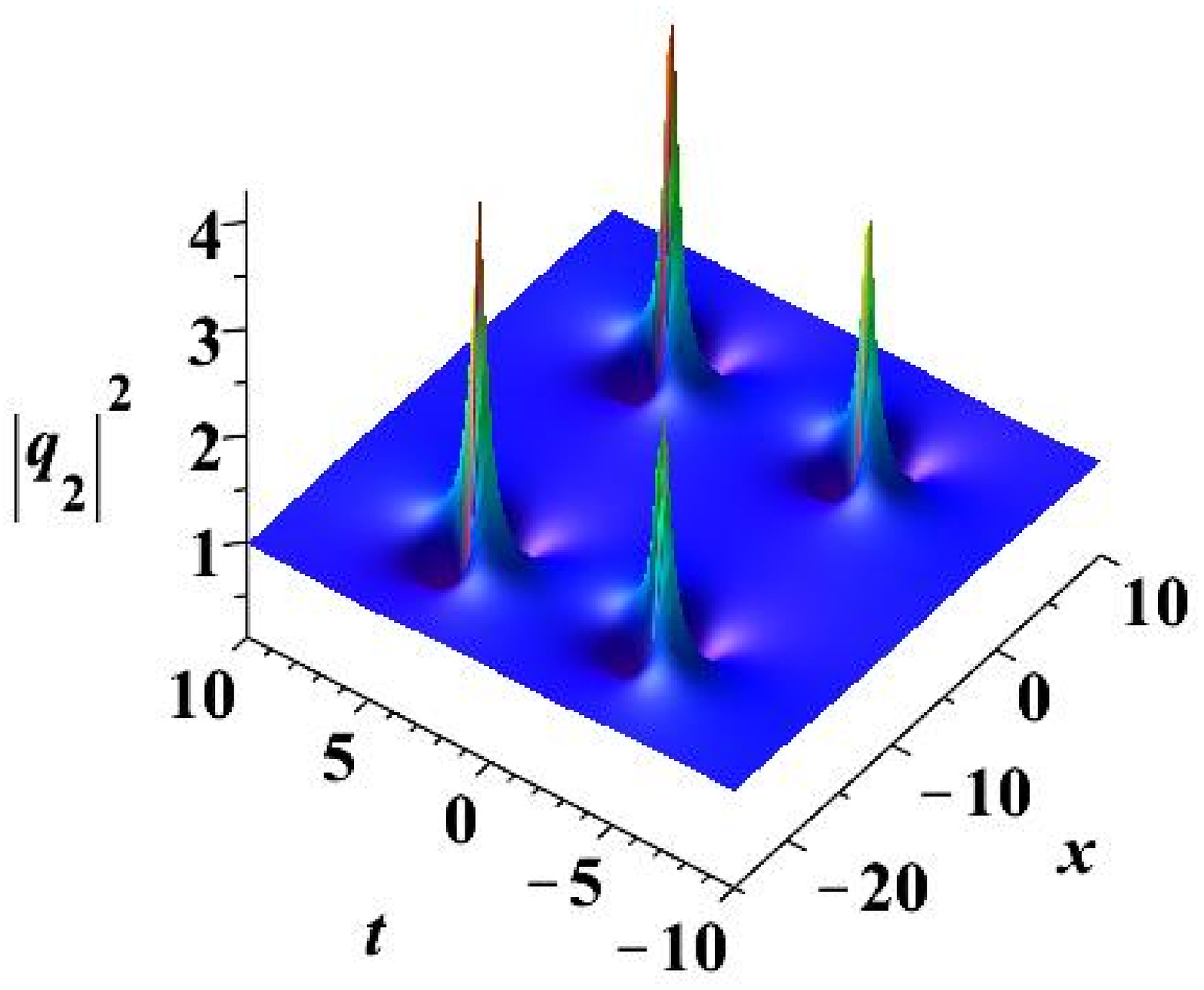}}
\caption{(color online) (a,b) The ``rhombus" structure for the second-order vector RW which contains four fundamental ones. The parameters are $f_1 = 0, f_2 = 0, g_1 = 1, g_2 = 0, h_1 = 0, h_2 = 10000$ , (c,d) The ``rectangle" structure for the second-order vector RW which contains four fundamental ones. The parameters are $f_1 = 0, f_2 = 0, g_1 = 1, g_2 = 1000, h_1 = 10, h_2 = 0$
. }\label{fig1}
\end{figure}
 The compact solution formula \eqref{NRW} can be used to derive N-th order RW solution. With $N=1$, the first order vector RW can be derived directly, which agrees well with the ones in \cite{Ling3,zhao2}.
 We find the dynamics structure of high order RW in the coupled system are much more abundant than the ones in scalar systems \cite{Yang,Ling,He,Ling2,Akhmediev}. Even for the second-order RW, whose distributions processes many different structures, which are quite different from the second order scalar ones.
As an example, we exhibit the dynamics behavior of second order
vector RW solution. Since the expressions of high order RW solution
are quite complicated, we will present them elsewhere.

\emph{The dynamics of second order vector rogue waves}--- There are
six free parameters in the generalized second order rogue wave
solution, denoted by $f_j$, $g_j$ and $h_j$ ($j=1, 2$), which can be
used to obtain different types or patterns for the rogue wave
dynamics.  We find that there are mainly two kinds of rogue wave
solutions which correspond to four fundamental RWs and six
fundamental ones obtained by setting $f_1=0$ and $f_1\neq 0$
respectively.
\begin{figure}[htb]
\centering
\subfigure[]{\includegraphics[height=36mm,width=42.5mm]{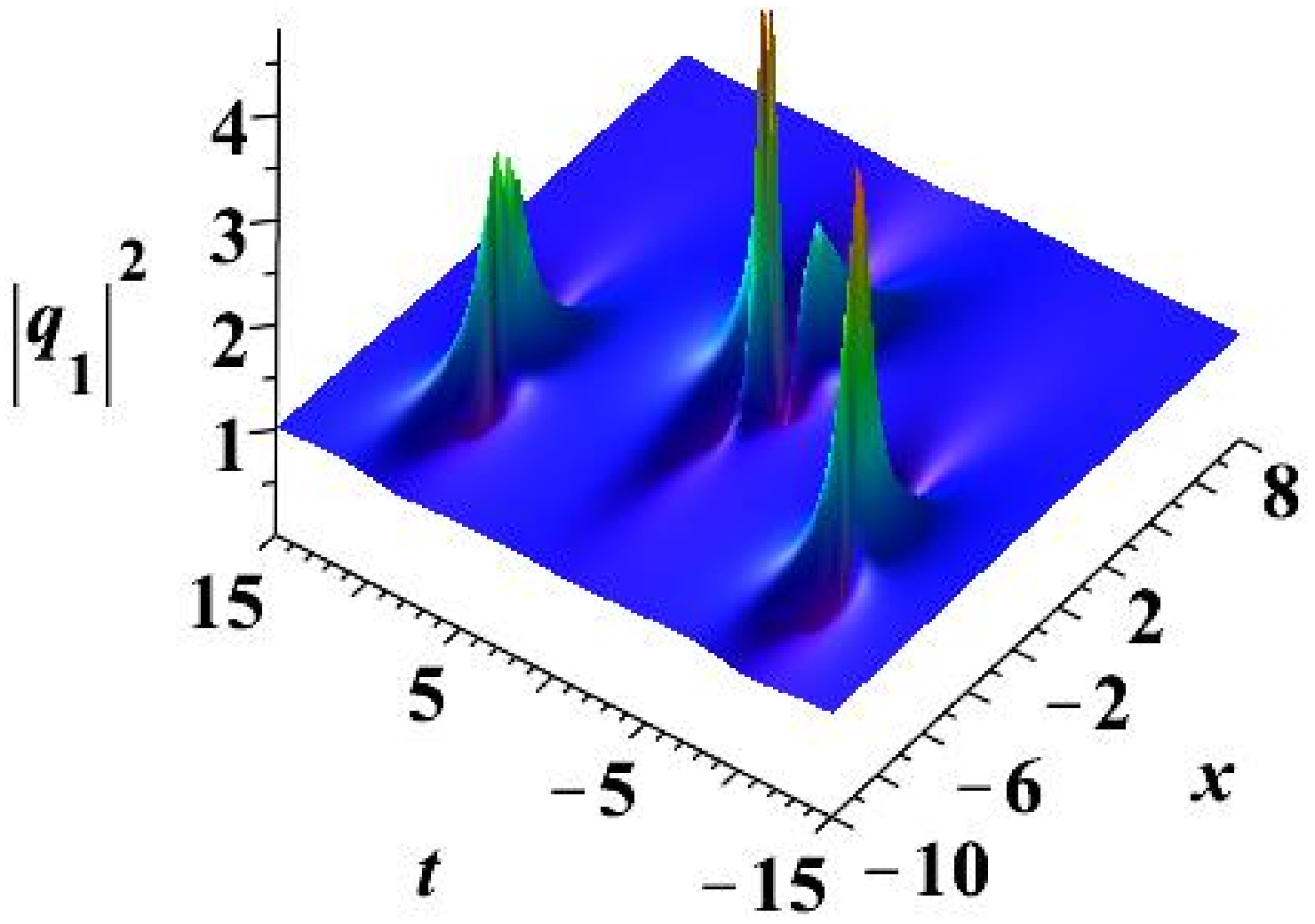}}
\hfil
\subfigure[]{\includegraphics[height=36mm,width=42.5mm]{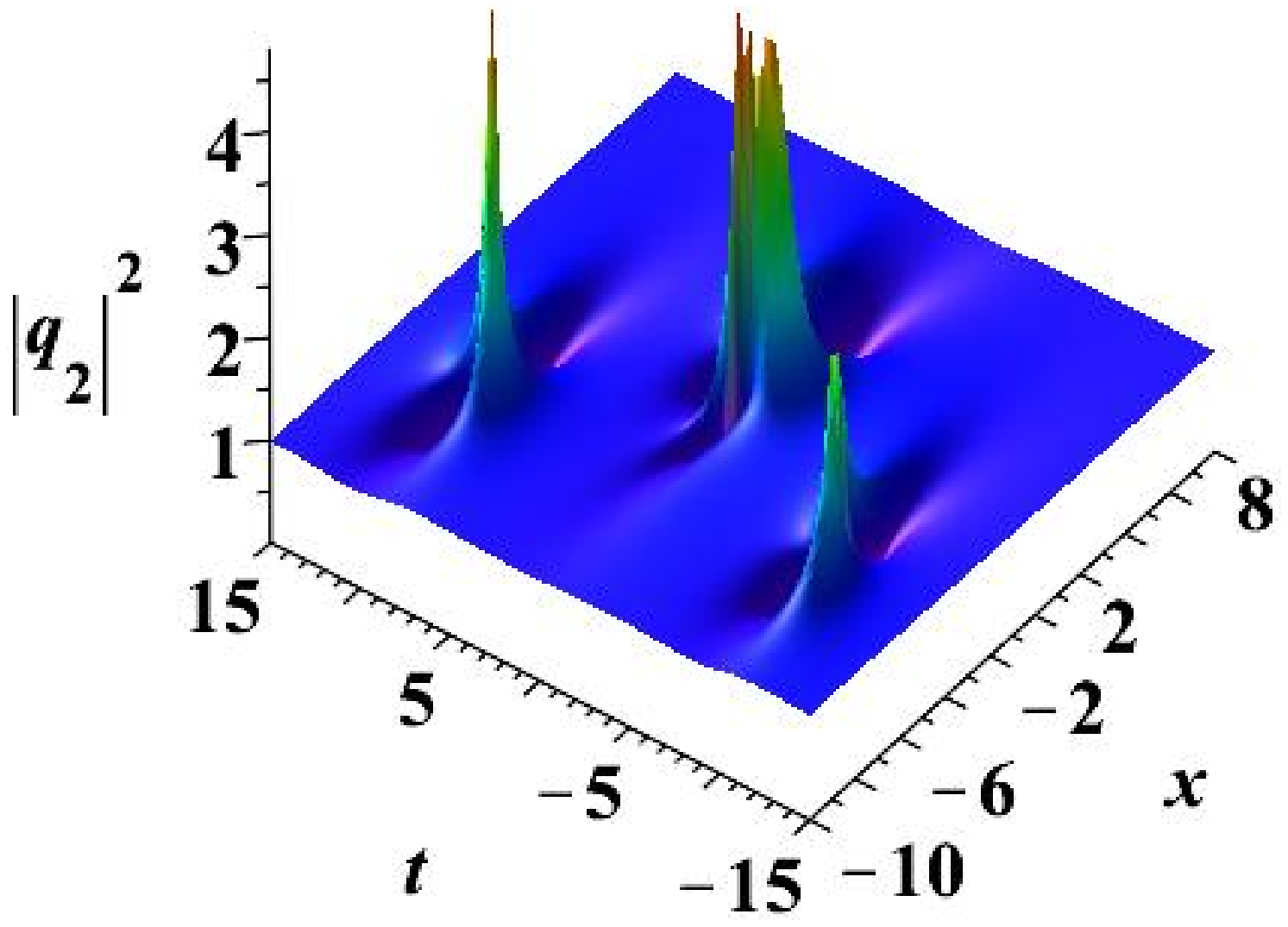}}
\hfil
\subfigure[]{\includegraphics[height=36mm,width=42.5mm]{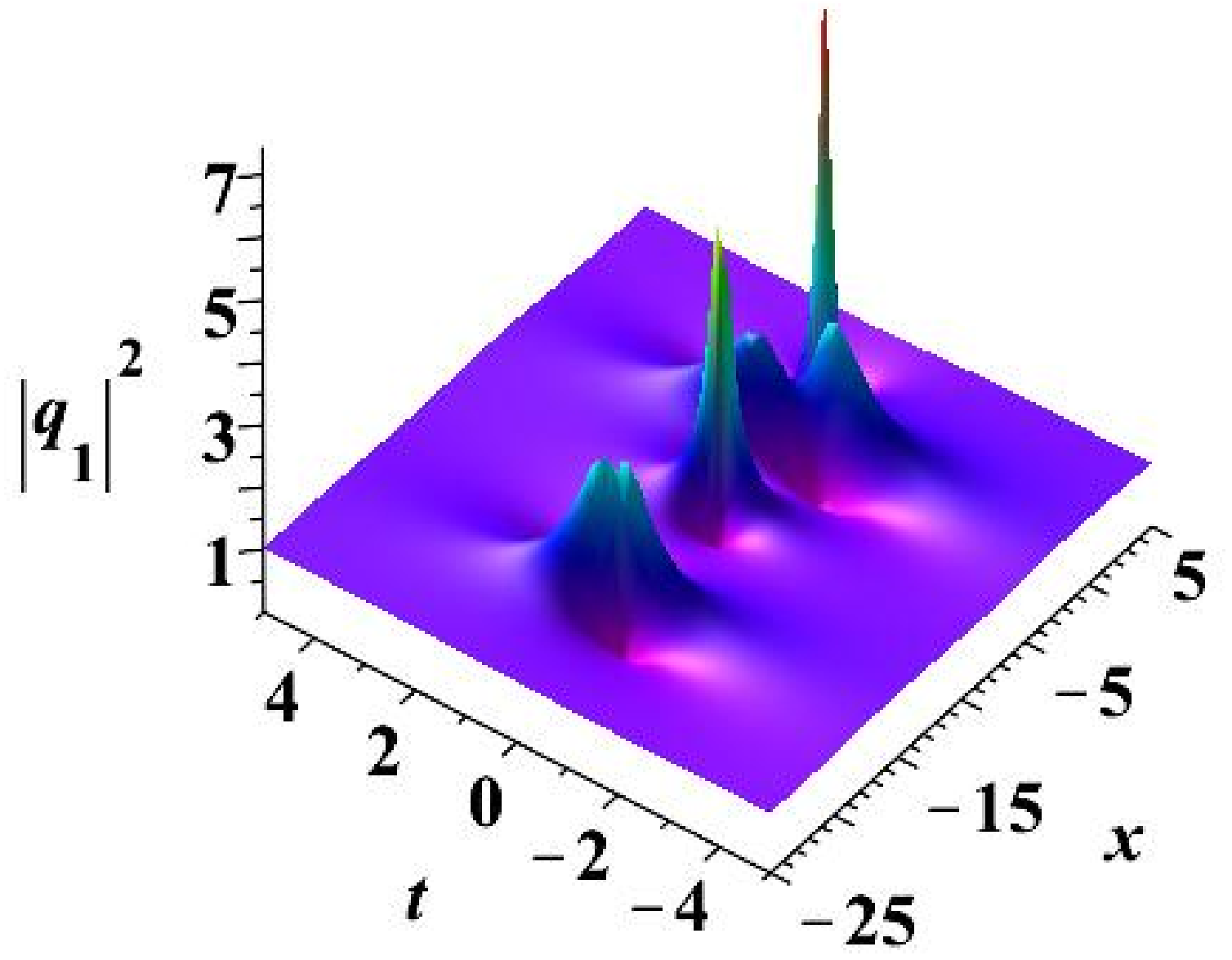}}
\hfil
\subfigure[]{\includegraphics[height=36mm,width=42.5mm]{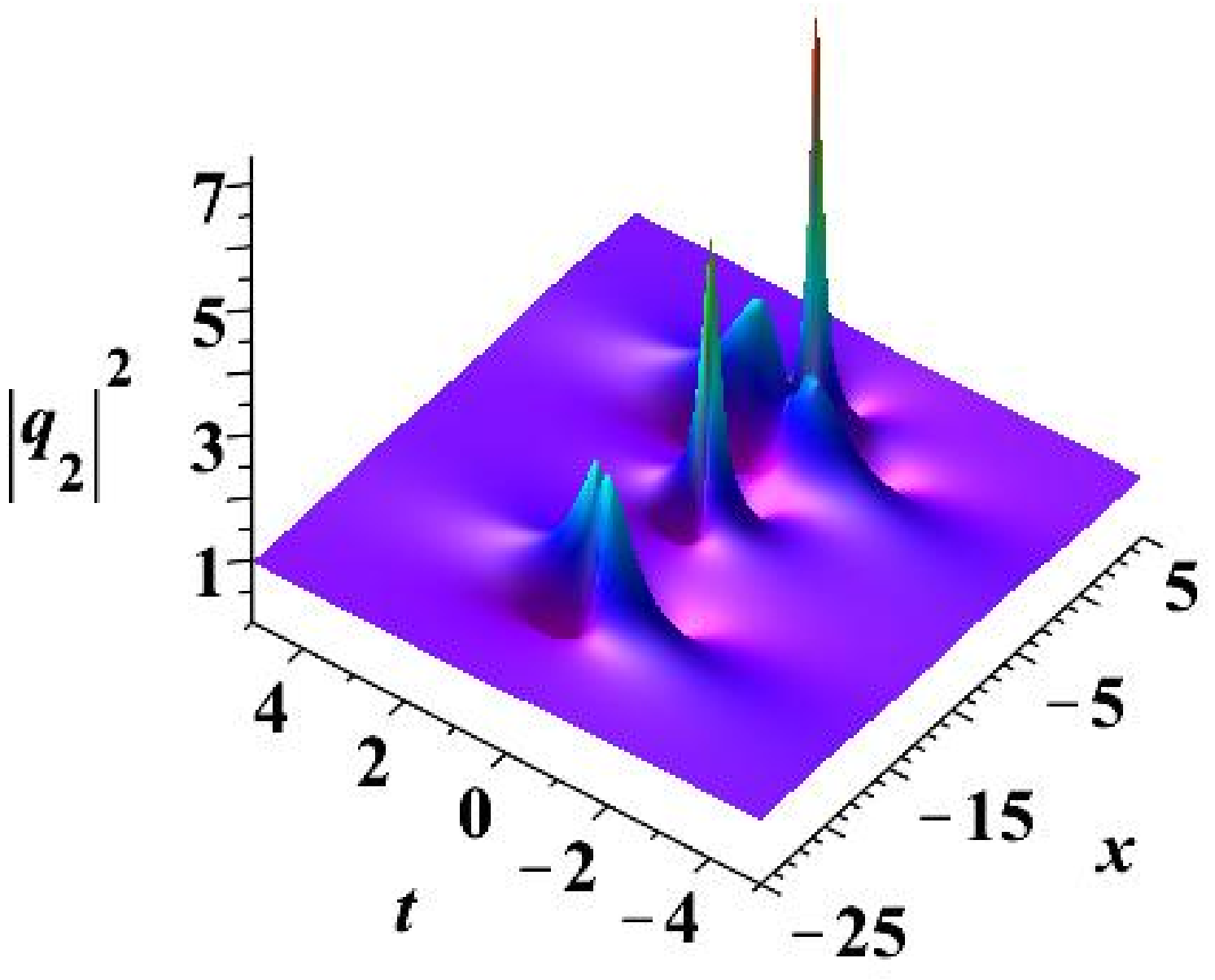}}
\caption{(color online) (a,b) The ``triangle" structure for the second-order vector RW which contains four fundamental ones. The parameters are $f_1 = 0, f_2 = 100, g_1 = 1, g_2 = 0, h_1 = 0, h_2 = 0$, (c,d) The ``line" structure for the second-order vector RW which contains four fundamental ones. The parameters are $f_1 = 0, f_2 = 0, g_1 = 1, g_2 = 0, h_1 = 10, h_2 = 0$.}\label{fig2}
\end{figure}
Firstly, we discuss the second order RW solution which
 possesses four fundamental RWs. The pattern is quite different from the ones in scalar NLS systems \cite{Ling2,Akhmediev}, for which it is impossible for four fundamental RWs to emerge on the temporal-spatial distribution plane.  To obtain this kind of solution, we merely need to choose
parameter $f_1=0$. We classify them by parameters $f_2,g_1, g_2, h_1, h_2$ whether or not are zeros. By this classification, there could be $2^5$ kinds of different solution which correspond to different patterns on the temporal-spatial distribution. In fact, the solution with different parameters here correspond to different ideal initial excitation for RW experimentally.  We  find there are mainly three types of the patterns, such as quadrilateral, triangle, and line structures.

The explicit shape of the quadrilateral can be varied though the parameters. As an example, we show two cases for the quadrilateral structure in Fig. 1.  The first case: the four RWs arrange with the ``rhombus" structure (Fig. \ref{fig1}(a,b)). The spatial-temporal distribution are similar globally in the two components, but the RW with highest peak emerge at different time, it appear at time $t=-5$ for the component $q_1$; and at $t=5$ for the component $q_2$. The second case: the four RWs arrange with ``rectangle" structure (Fig. \ref{fig1}(c,d)). It is seen that the peak values of two RWs on the right hand is much higher than the ones on the left hand in the component $q_1$. The character is inverse for the component $q_2$.

Varying the other parameters, we can observe the interaction between the four RWs. When two of them fuse into be a new RW, the three RWs can emerge with ``triangle" structure on the temporal-spatial distribution (Fig. \ref{fig2}(a,b)). The structure of the triangle can be changed by vary the parameters. Especially, the three RWs can emerge in a ``line" (Fig. \ref{fig2}(c,d)),
 which is perpendicular with the $t$ axes. Namely, at a certain time, three or four RWs can emerge synchronously.
\begin{figure}[htb]
\centering
\subfigure[]{\includegraphics[height=36mm,width=42.5mm]{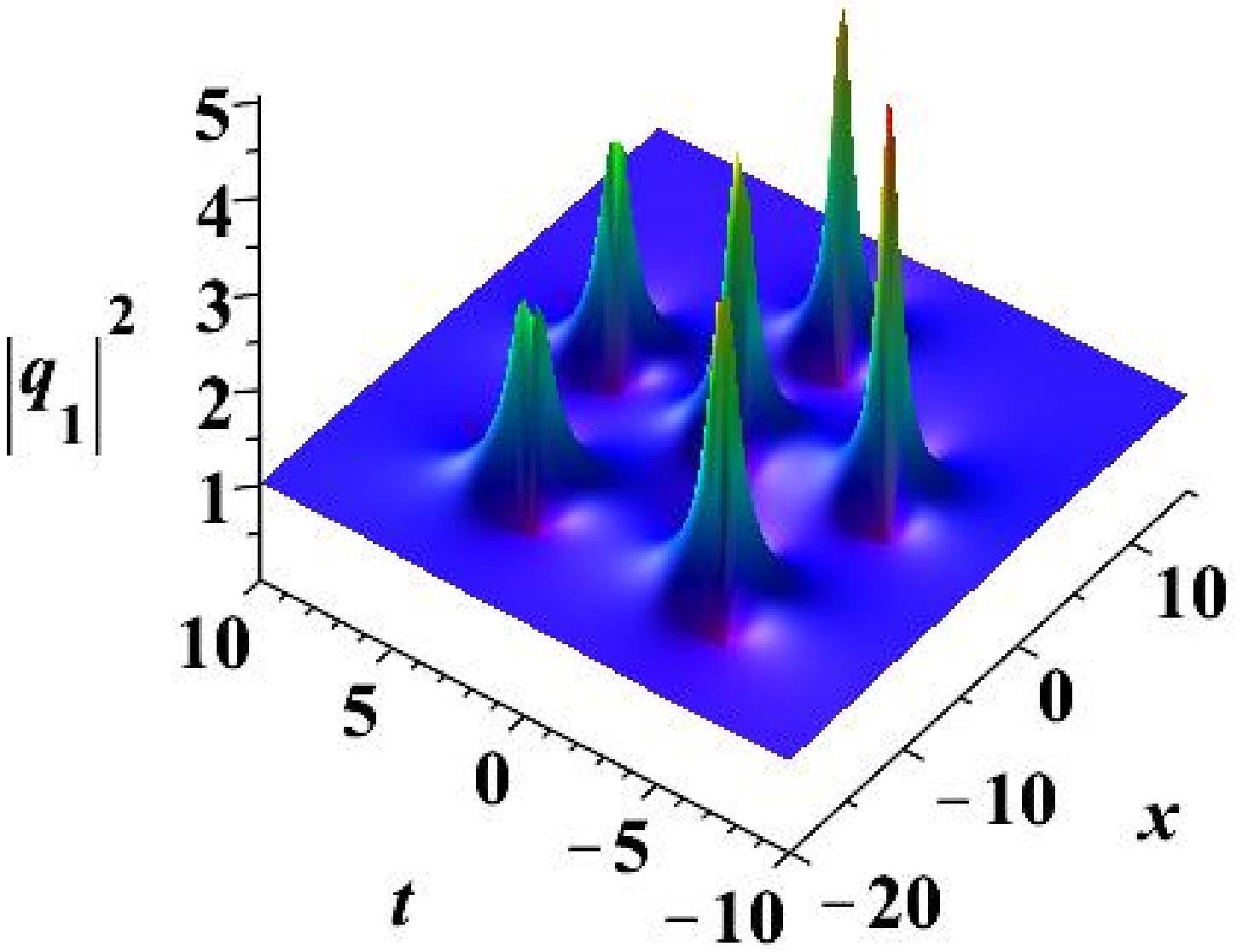}}
\hfil
\subfigure[]{\includegraphics[height=36mm,width=42.5mm]{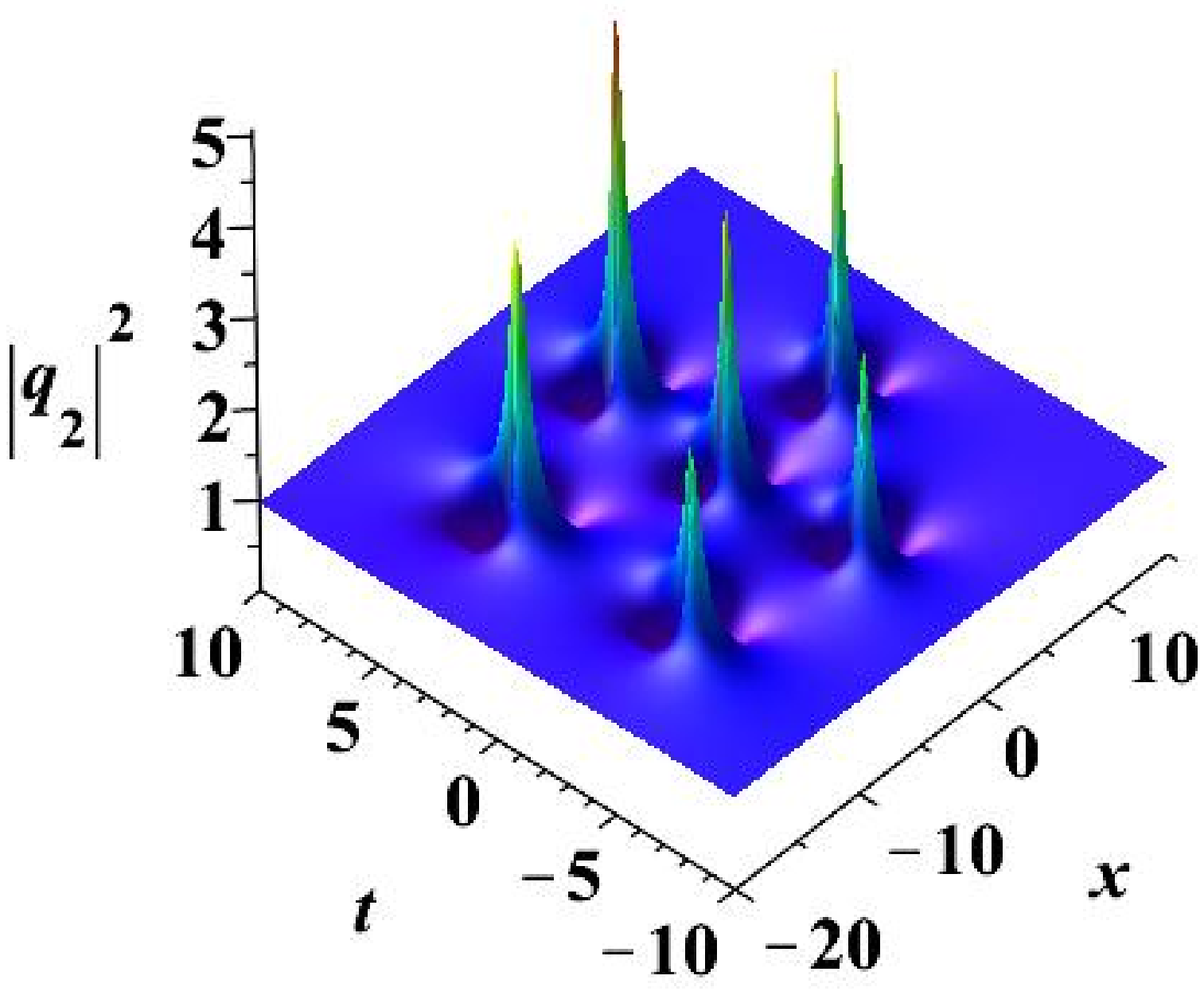}}
\hfil
\subfigure[]{\includegraphics[height=36mm,width=42.5mm]{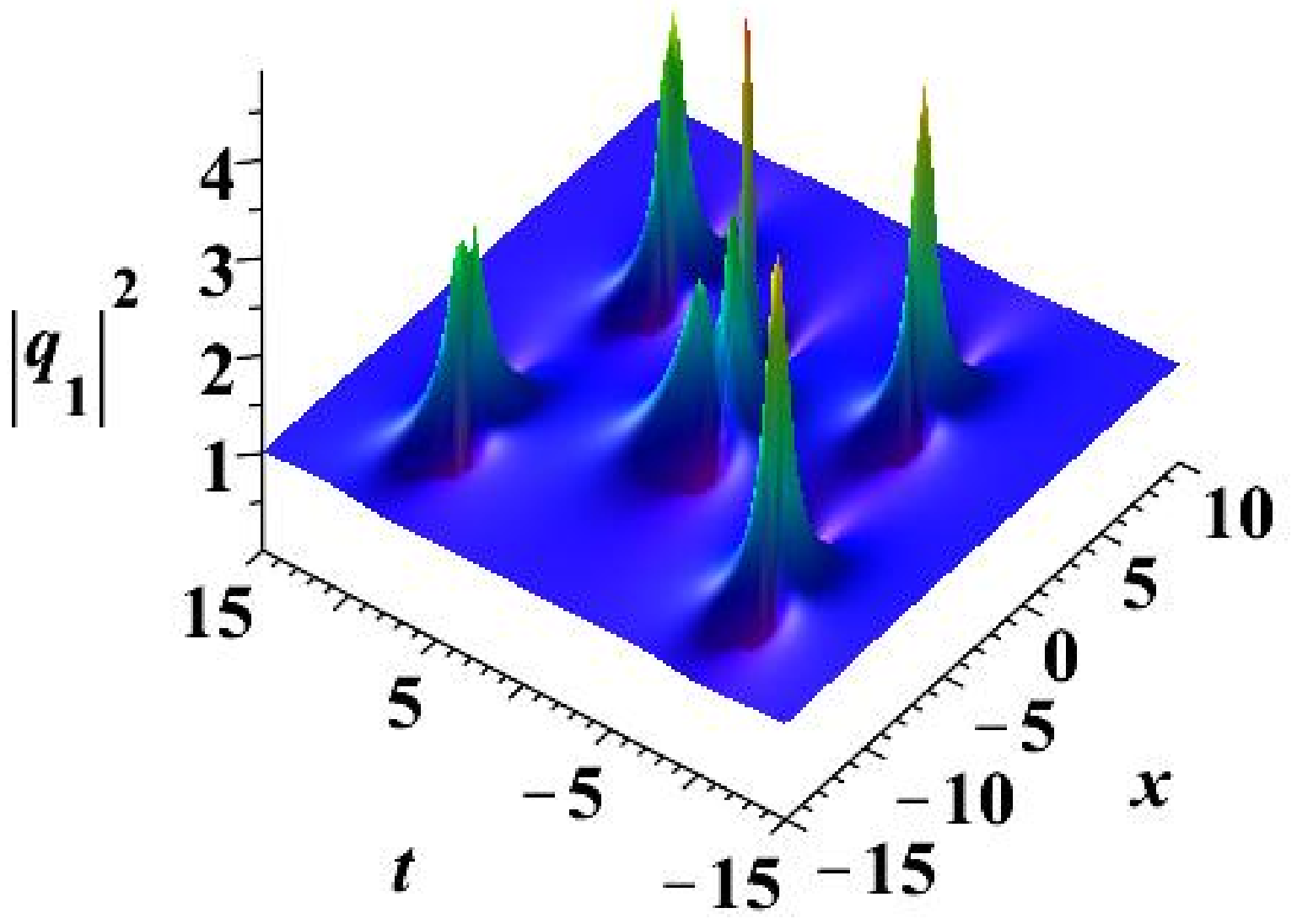}}
\hfil
\subfigure[]{\includegraphics[height=36mm,width=42.5mm]{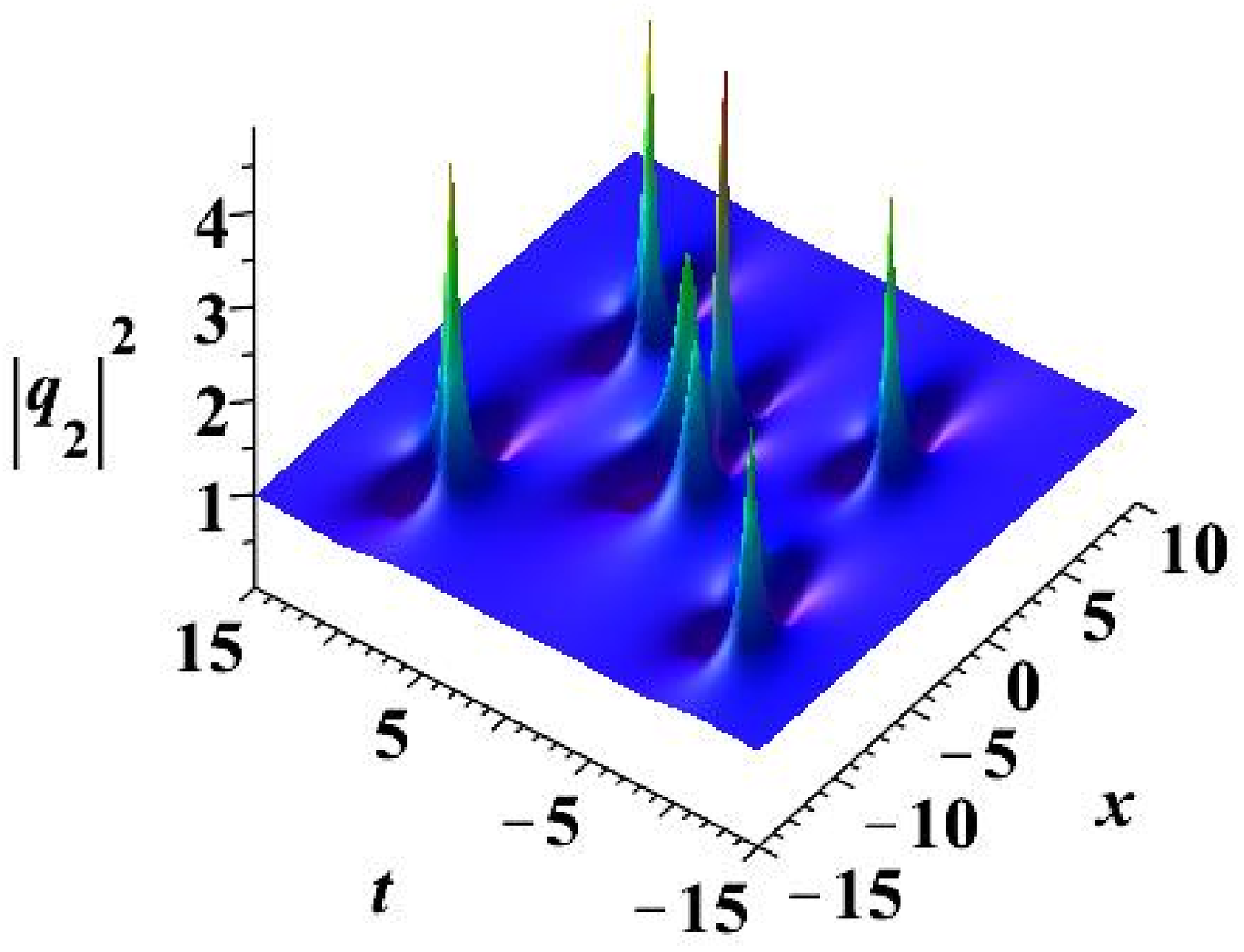}}
\caption{(color online) (a,b)The ``pentagon" structure for the second-order vector RW which contains six fundamental ones. The parameters are $f_1 = 1, f_2 = 0, g_1 = 0, g_2 = 0, h_1 = 0, h_2 = 10000$ , (c,d) The ``rectangle" structure for the second-order vector RW which contains six fundamental ones. The parameters are $f_1 = 1, f_2 = 0, g_1 = 0, g_2 = 0, h_1 = 100, h_2 = 0$.}\label{fig3}
\end{figure}

Secondly, we consider the second case of second order RW solutions,
which possess six fundamental RWs. To obtain this kind of solution,
we choose the parameter $f_1=1$. One can simply classify this
solution $2^{5}$ types by parameters $f_2,g_1,g_2,h_1,h_2$. But the
six fundamental RWs can constitute many different structures, such
as ``pentagon", ``quadrilateral", ``triangle", and ``line"
structures. As an example, we show the
``pentagon" structure in Fig. \ref{fig3}(a,b). The pentagon structure
can be varied too though changing the parameters. There is one RW in
the internal region of the pentagon, and its location in the
distribution plane can be varied too. This case is similar to the
``pentagon" structure of the third order RW of scalar NLS equation
\cite{Ling2,Akhmediev}. The six RWs can be arranged with the
``rectangle" structure through varying the parameters, such as the
one in Fig. \ref{fig3}(c,d). The structure is similar to the one in
Fig. \ref{fig1}(c,d). But there is a new RW insider ``rectangle",
which is formed by the interaction of the other two fundamental RWs.

\begin{figure}[htb]
\centering
\subfigure[]{\includegraphics[height=36mm,width=42.5mm]{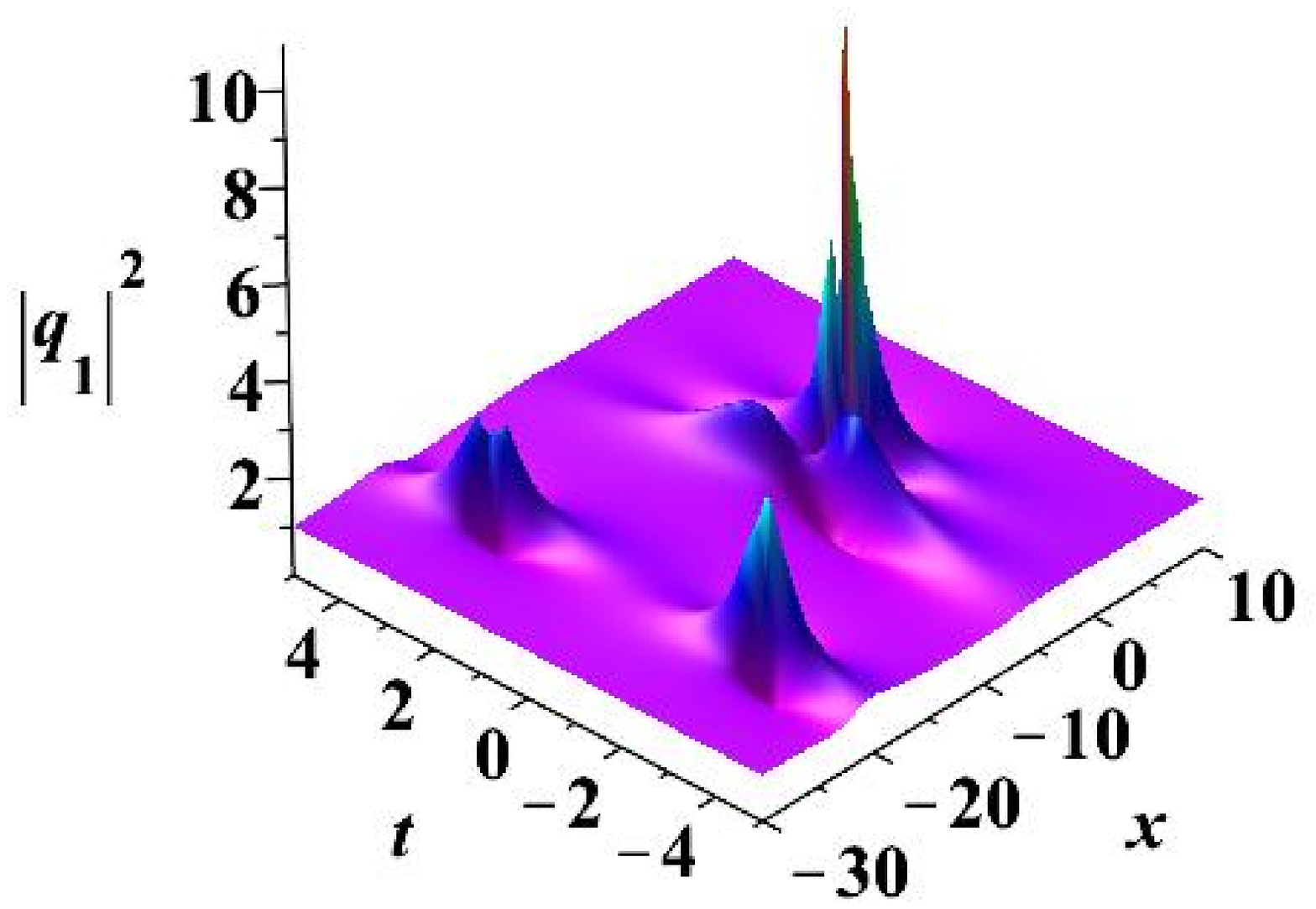}}
\hfil
\subfigure[]{\includegraphics[height=36mm,width=42.5mm]{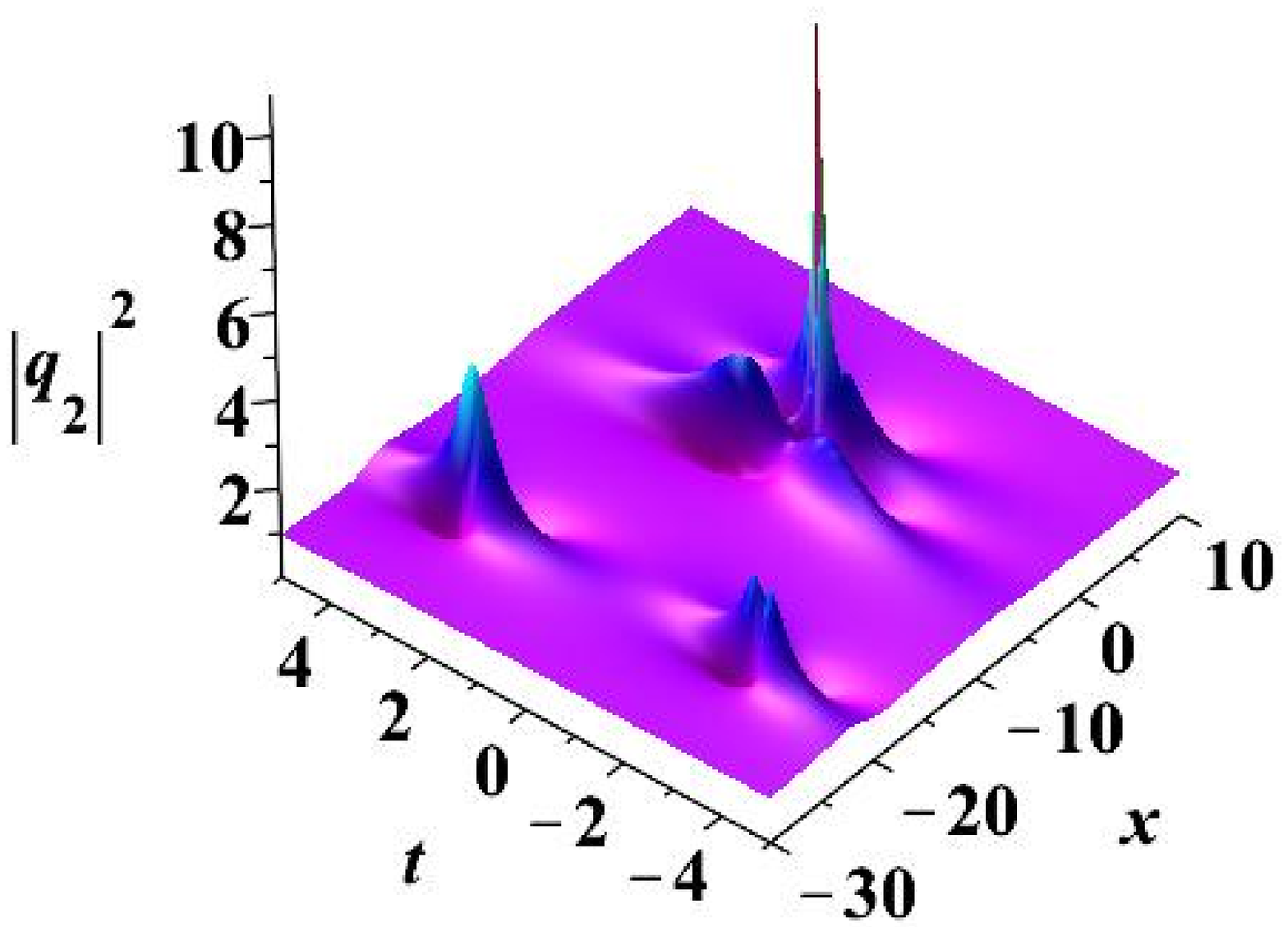}}
\hfil
\subfigure[]{\includegraphics[height=36mm,width=42.5mm]{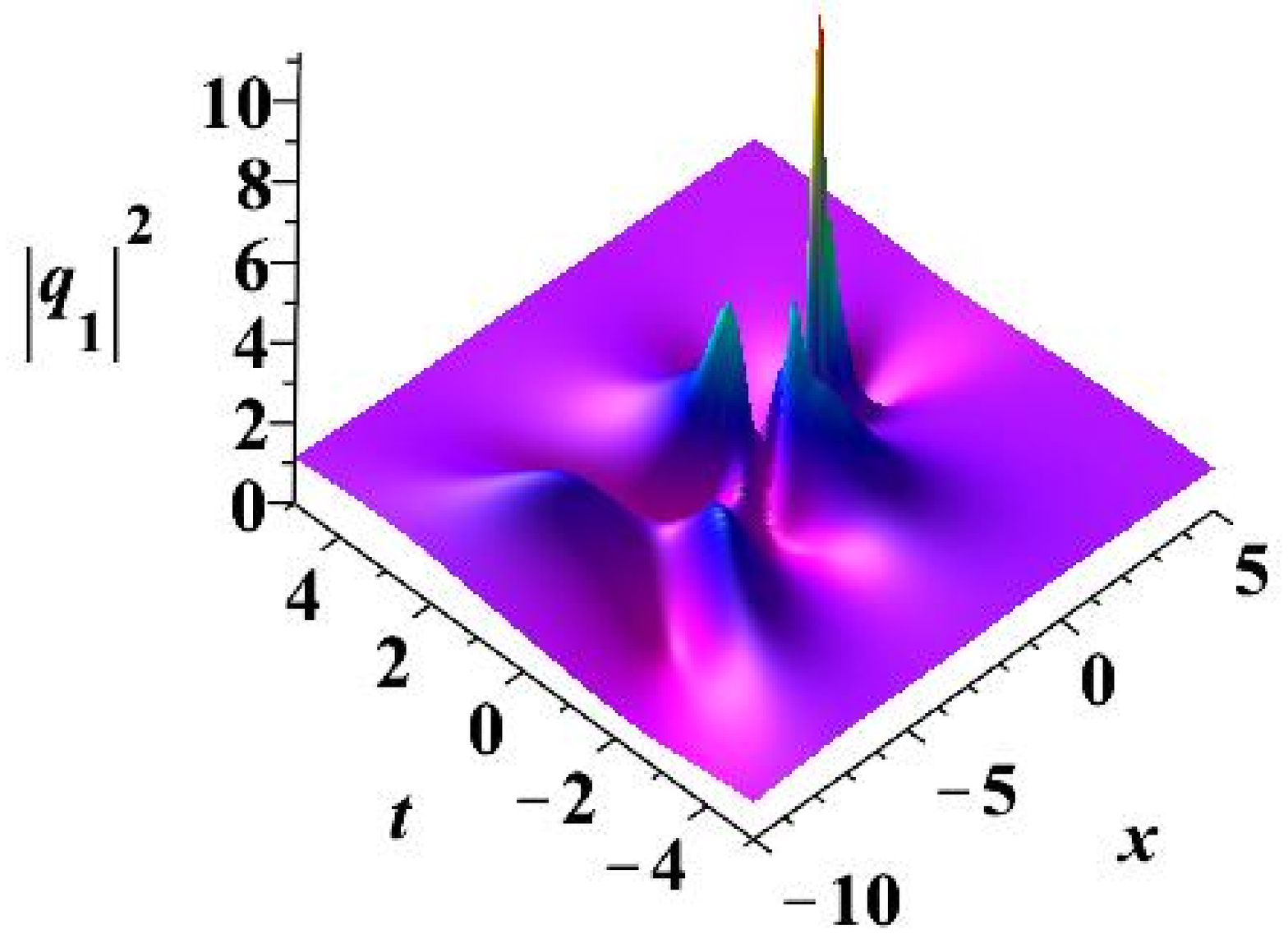}}
\hfil
\subfigure[]{\includegraphics[height=36mm,width=42.5mm]{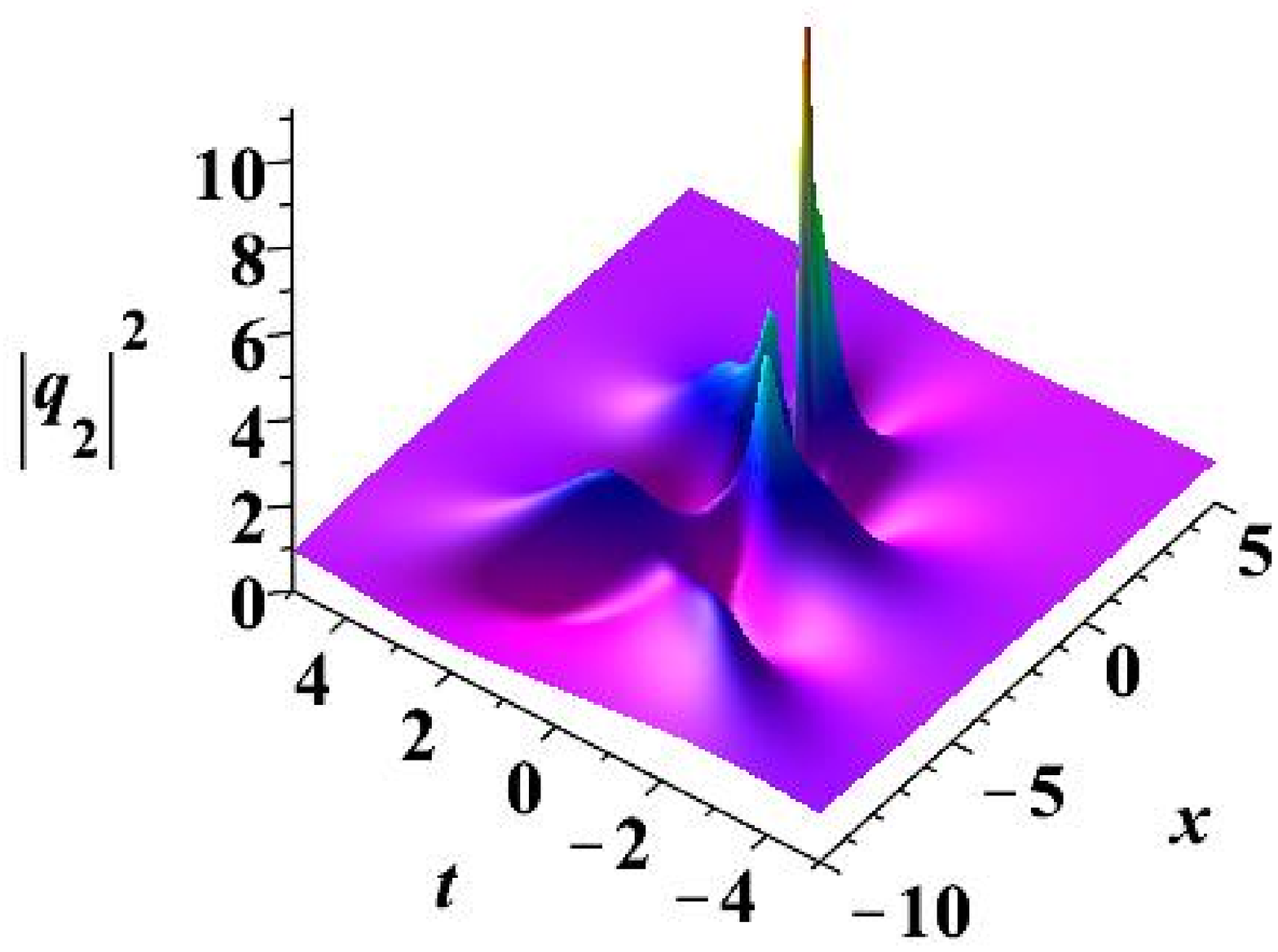}}
\caption{(color online) (a,b) The ``triangle" structure for the second-order vector RW which contains six fundamental ones. The parameters are $f_1 = 1, f_2 = 0, g_1 = 10, g_2 = 0, h_1 = 0, h_2 = 0$
.  (c,d) The ``line" structure for the second-order vector RW which contains six fundamental ones. The parameters are $f_1 = 1, f_2 = 0, g_1 = 0, g_2 = 0, h_1 = 0, h_2 = 0$.  }\label{fig4}
\end{figure}

Similar to the ones for four fundamental RWs case, the six RWs can
be arranged with the ``triangle" structure too, such as the ones in
Fig. \ref{fig4}(a,b). In this case, there are two fundamental RWs
and a new RW to form a triangle. The new RW is formed by the
interaction between the other four fundamental RWs. Moreover, the
RWs can be arranged with ``line" structure too, shown in Fig.
\ref{fig4}(c,d). There should be six fundamental RWs arranged in one
line case, but it is very complicated to derive the case since the
parameters are too many to be managed well. We just show a
particular one of the ``line" cases.

\emph{Possibilities to observe these vector rogue waves}---
Considering the experiments on RW in nonlinear fibers with anomalous
GVD \cite{Kibler,Solli}, which have shown that the simple
scalar NLS could describe nonlinear waves in nonlinear fibers well,
we expect that these different vector nonlinear waves could be
observed in two-mode nonlinear fibers. Consider the case that the operation wavelength of each mode
is nearly $1.55 \ \mu m$, the GVD coefficients are $-20\ ps^2  km^{-1}$
in the anomalous regime, and the Kerr coefficients are nearly
$1.1\ W^{-1} km^{-1}$, corresponding to the self-focusing effect in the
fiber \cite{Dudley2}. The unit in $x$ direction will be denoted as $0.23\ ps$,
and the one in $t$ will be denoted as $0.55\ km$. We discuss the proper conditions
to observe the vector RWs. One can introduce two distinct modes to the
nonlinear fiber operating in the anomalous GVD regime
\cite{Afanasyev,Ueda}. The spontaneous
development of RW seeded from some perturbation should be on the
continuous waves as the ones in \cite{Kibler,Solli}.
The continuous wave background intensities in the two modes should be equal (assume to be $30\ W$). The frequency difference between the two mode should be $0.23\ ps^{-1}$. Then the ideal initial
optical shape can be given by the exact second-order vector RW solution with explicit coefficients in the scaled units. As shown in \cite{Kibler}, the Peregrine
RW characteristics can appear with initial conditions that do not
correspond to the mathematical ideal, the vector RWs could be observed in the nonlinear fiber with approaching the ideal initial excitation form.

\emph{Conclusions}--- We find that there are mainly two kinds of
rogue wave solutions for the second order vector RW in two-component
coupled NLSE, which correspond to four fundamental RWs and six
fundamental ones obtained by setting $f_1=0$ and $f_1\neq 0$
respectively. The distribution patten
for vector RWs are much abundant than the ones for scalar rogue
waves. These results could be helpful to understand the
complex RW phenomena.


\end{document}